\documentclass[preprint,showpacs,preprintnumbers,amsmath,amssymb]{revtex4}
\usepackage{graphicx}
\usepackage{dcolumn}
\usepackage{bm}
\begin{document}
\newcommand{\be}{\begin{equation}}
\newcommand{\ee}{\end{equation}}
\newcommand{\bea}{\begin{eqnarray}}
\newcommand{\eea}{\end{eqnarray}}
\newcommand{\non}{\nonumber}
\title{A momentum-space representation of Green's functions with modified dispersion relations on general backgrounds}
\author{Massimiliano Rinaldi}
\affiliation{D\'epartement de Physique Th\'eorique, Universit\'e de Gen\`eve, \\
24, quai E. Ansermet 1211 Gen\`eve 4, Switzerland.} 
\date{\today}
\begin{abstract}
\noindent We consider the problem of calculating the Green's functions associated to a massive scalar field with modified dispersion relations. We analyze the case when dispersion is modified by higher derivative spatial operators acting on the field orthogonally to a preferred direction, determined by a unit time-like vector field. By assuming that the integral curves of the vector field are geodesics, we expand the modified Klein-Gordon equation in Fermi normal coordinates. By means of a Fourier transform, we find a series representation in momentum-space of the Green's functions. The coefficients of the series are geometrical terms containing combinations of the Ricci tensor and the vector field, as expected from previous calculations with different methods and for specific backgrounds.

\end{abstract}


\maketitle

\section{Introduction}

\noindent In recent times, there has been a considerable amount of investigations on the trans-Planckian problem. As a satisfactory theory of quantum gravity is not available yet, some researchers have focussed on the possibility that the unknown physics beyond the Plank scale appears in the form of a modification of the dispersion relations above a certain energy-scale. The consequences of this assumption have been widely examined in relation to the inflationary cosmology \cite{brand,lubo}, the Hawking radiation \cite{bhole}, and the Unruh effect \cite{unruheff}.  In some cases, modified dispersion relations (MDR) locally break the Lorentz-invariance (for a review, see \cite{Matt}). However, one can keep the general covariance of the action, by assuming the existence of a preferred frame, encoded by a unit time-like vector field, which breaks dynamically the Lorentz invariance \cite{Jacob}. In this setup, often named ``aether theory", the vector field is independent of the metric, in the sense that it represents a new degree of freedom in the Einstein-Hilbert action. In the most general case, the action contains several new terms associated to the vector field, its covariant derivatives and contractions with the Ricci tensor. All these terms carry coefficients which are constrained by observations, and recent results show that the aether theory is not incompatible with available data (see, e.g., \cite{ruth}). 

The vector field can be coupled to covariant derivatives of quantum fields, thus generating modified dispersion relations. In flat space, the propagator associated to these fields contains higher powers of the spatial momenta, which displays explicitly the breaking of the Lorentz group down to a spatial rotation subgroup. Thus, also in curved space one expects that the Hadamard form of the propagator is modified. As a consequence, the fundamental structure of quantum observables is certainly different with respect to the Lorentz-invariant case. As, in the semi-classical theory, the expectation value of the renormalized stress tensor is the source term of the Einstein equations (see for instance \cite{BD}), the calculation of the modified  Green's functions is of fundamental importance in order to establish the quantum back-reaction, and, hence, the effects of trans-Planckian modes. Some investigation on this issue has recently begun, especially in the context of cosmology. For example, the renormalized stress tensor has been calculated, via adiabatic regularization, in the cases of FLRW and Bianchi I cosmological backgrounds \cite{mazz1,mazz2}. However, it would be desirable to have a general expression, which can be applied to any background. 

The aim of this paper is to find a general expression for the Green's functions with modified dispersion relations. In a previous paper \cite{io}, we found a representation of these functions in momentum-space, in the case when the background is ultra-static. Here, we generalize this result for any background and for any analytic (in momentum-space) dispersion relation. When the latter is Lorentz-invariant, there are various equivalent techniques to calculate the Green's functions, such as the deWitt-Schwinger expansion in coordinate space \cite{deWitt}, and the Bunch and Parker expansion in momentum-space \cite{BP}. When MDR are present, however, the first technique is difficult to apply as it heavily relies upon Lorentz-invariance, while the second does not need this symmetry. In fact, Lorentz-invariance in the Bunch and Parker expansion allows to reduce the number of terms in the expansion, but it is not essential, as already shown for the ultra-static case in \cite{io}. For general backgrounds, however, the Bunch and Parker method leads to algebraically complicated results because of the Riemann normal coordinates (RNC) used in the expansion of the operators. This coordinate system is very useful when one needs to expand the metric around one fixed point. But when, besides the metric, there exists also an independent vector field, together with differential operators associated to it, the expansion of the latter becomes very complicated in RNC. As the vector field define a congruence of time-like curves, it would be more convenient to find a coordinate system adapted to these. 

Such a system exists and it was discovered by Fermi \cite{fermi}, and  developed later by several authors (see, e.g. \cite{Misner, FNC}). It generalizes the RNC, in the sense that it allows to define a local orthonormal coordinate system non only about one fixed point, but  about an entire geodesic curve \cite{poisson}. Thus, the use of the so-called Fermi normal coordinate (FNC) allows to simplify the problem of expanding the operators associated to the vector field, and provides for a well-defined orthonormal frame. With this frame, as we will show, it is possible to construct a momentum-space representation of the Green's function by Fourier transforming the Klein-Gordon operator. 

The plan of the paper is the following. In Section II, we introduce the modified Klein-Gordon operator in the context of the aether theory. In Section III, we calculate the momentum-space representation of the Green's functions. Most of the technicalities are confined in the appendices: in the first, we briefly sketch the FNC construction, while, in the second, we display the expansion in FNC of the differential operators contained in the Klein-Gordon equation. We conclude, in Section IV, with a discussion of the results and few remarks.
In this paper, we set $\hbar=c=1$, and the signature is $(-,+,+,\ldots)$.

\section{Aether theory and MDR}

\noindent In the aether theory, one considers the Einstein-Hilbert action implemented by terms depending on a unit time-like vector field $u^{\mu}$. The general form in $N$ dimensions is given by
\bea
S={1\over 16\pi G}\int d^Nx\sqrt{-g}\left(R-2\Lambda+{\cal L}_u\right)\ ,
\eea
where $R$, $\Lambda$, and $G$ are the Ricci scalar, the cosmological constant and the Newton constant respectively. The term ${\cal L}_u$ is the Lagrangian associated to the vector field, and its general form reads \cite{Jacob}
\bea\label{lag}
{\cal L}_u=-\lambda(u^{\mu}u_{\mu}+1)-b_1F_{\mu\nu}F^{\mu\nu}-b_2(\nabla_{\mu}u^{\mu})^2-b_3R_{\mu\nu}u^{\mu}u^{\nu}-b_4u^{\alpha}u^{\beta}\nabla_{\alpha}u_{\mu}\nabla_{\beta}u^{\mu}\ .
\eea
In this expression, $F_{\mu\nu}=2\nabla_{[\mu}u_{\nu]}\ $, and the constant $\lambda$ acts as a Lagrange multiplier, which  ensures that the vector field is unit and time-like. Finally, $R_{\mu\nu}$ is the Ricci tensor, and the coefficients $b_1\ldots b_4$ are arbitrary constant. The form of this Lagrangian guarantees that the theory is generally covariant, even though the vector field dynamically breaks local Lorentz invariance, as it determines a preferred direction. Some of the terms in Eq.\ (\ref{lag}) vanish whenever the integral curves of the vector field are geodesics. In such a case, in fact, the vector field is everywhere hypersurface orthogonal and $F_{\mu\nu}=0$. Also, the acceleration $a^{\mu}=u^{\alpha}\nabla_{\alpha}u^{\mu}=0$, and the term proportional to $b_4$ vanishes. In the following, we assume that the integral curves are geodesics, thus the observer comoving with $u^{\mu}$ is free-falling. In any case, to analyze the equations of motion, it is always convenient to choose a background of the form
\bea
g_{\mu\nu}dx^{\mu}dx^{\nu}=-(u_{\alpha}dx^{\alpha})^2+q_{\mu\nu}dx^{\mu}dx^{\nu}\ ,
\eea
where $q_{\mu\nu}=g_{\mu\nu}+u_{\mu}u_{\nu}$ is the induced metric on the hypersurfaces orthogonal to $u^{\mu}$ \cite{lubo,Jacob}. We now consider a free, massive, minimally coupled scalar field with a Lagrangian given by
\bea\label{philag}
{\cal L}_{\phi}=-{1\over 2}\left[g^{\mu\nu}\partial_{\mu}\phi\partial_{\nu}\phi+m^2\phi^2+\sum_{p,q}A_{pq}\left({\cal D}^{2p}\phi\right)\left({\cal D}^{2q}\phi\right)\right]\ ,
\eea
where the operator ${\cal D}$ is defined such that
\bea
{\cal D}^2\phi=q^{\alpha}{}_{\mu}\nabla_{\alpha}(q^{\mu}{}_{\beta}\nabla^{\beta}\phi)\ ,
\eea
and the coefficients $A_{pq}$ are arbitrary \cite{lubo,Jacob}. This Lagrangian explicitly breaks the local Lorentz group to a rotational subgroup \cite{Jacob}. The Klein-Gordon operator associated to the field can be written in the form
\bea
(\square - m^2)\phi+F\left[{\cal D}^2\right]\phi=0\ ,
\eea
where $F$ is some functional of the operator ${\cal D}^2$, such that it can formally be expanded as 
\bea\label{Fexp}
F\left[{\cal D}^2\right]\phi=\sum_{n=2}^{\infty}\alpha_{2n}{\cal D}^{2n}\phi\ ,
\eea 
with arbitrary coefficients $\alpha_{2n}$. This operator is responsible for the modification of the dispersion relation as, in momentum-space, it introduces higher powers of $p^2$, the square of the spatial momentum. For our purposes, it is more convenient to write $\square \phi$ as \cite{lubo}
\bea\label{Dsquare}
\square \phi= {\cal D}^2\phi-u^{\alpha}u^{\beta}\nabla_{\alpha}\nabla_{\beta}\phi-K\, u^{\alpha}\partial_{\alpha}\phi\ ,
\eea 
where $K=q^{\mu\nu}\nabla_{\mu}u_{\nu}$ is the trace of the extrinsic curvature of the hypersurface orthogonal to $u^{\mu}$. In this way, we can write the Klein-Gordon equation as
\bea
{\cal D}^2\phi-u^{\alpha}u^{\beta}\nabla_{\alpha}\nabla_{\beta}\phi-K\, u^{\alpha}\partial_{\alpha}\phi-m^2\phi+F\left[{\cal D}^2\right]\phi=0\ .
\eea
If $F$ can be expanded as in Eq.\ (\ref{Fexp}), we can include the first term of the above equation into the last one, and finally write the $N$-dimensional Klein-Gordon equation as
\bea\label{KG}
\sum_{n=1}^{\infty}\alpha_{2n}{\cal D}^{2n}\phi-u^{\alpha}u^{\beta}\nabla_{\alpha}\nabla_{\beta}\phi-K\, u^{\alpha}\partial_{\alpha}\phi-m^2\phi=0\ ,
\eea
with the requirement that $\alpha_2=1$, while the other $\alpha$'s are arbitrary.

\section{Momentum-space Representation of the Green's Functions}

\noindent In this section, we find an expansion up to second order of the metric in FNC, of the Green's functions $G(x,x')$ associated to the scalar field $\phi$ of Eq.\ (\ref{KG}). These must satisfy the equation
\bea
\left[\sum_{n=1}^{\infty}\alpha_{2n}{\cal D}^{2n}-u^{\alpha}u^{\beta}\nabla_{\alpha}\nabla_{\beta}-K\, u^{\alpha}\partial_{\alpha}-m^2\right]G(x,x')=-{1\over \sqrt{g}}\delta(x-x')\ ,
\eea
where $x$ and $x'$ are the coordinates of two nearby points, and the differential operators act, by convention, on $x$. In order to write this equation in terms of symmetric biscalars, we need to define \cite{BP,poisson-liv}
\bea\non
G(x,x')&=&g(x)^{-1/4}\bar G(x,x') g(x')^{-1/4}\\\non\\
g^{-1/2}(x)\delta(x-x')&=&g^{-1/4}(x)\delta(x-x')g^{-1/4}(x')\ .
\eea
If we choose the FNC as local coordinate system, we can identify $P$ as the point with coordinate $x'=(0,0,\ldots, 0)$, and $Q$ as the point with coordinate $x=(\tau,x^a)$, see Fig.\ (\ref{pic}) in Appendix I. Hence, the Green's functions $\bar G$ must satisfy the equation
\bea\label{Greeneq}
g^{1/4}\left[\sum_{n=1}^{\infty}\alpha_{2n}{\cal D}^{2n}-u^Au^B\nabla_A\nabla_B-K\, u^A\partial_A\right](g^{-1/4}\bar G)-m^2\bar G=-\delta(x^a)\delta(\tau)\ .
\eea
The convention is that  the indices $(A,B,\ldots)$ label space-time coordinates while $(a,b,\ldots)$ spatial coordinates only, see Appendix I. We now wish to expand this equation in FNC up to the second order. By using Eqs.\  (\ref{Hdet}) and (\ref{Dfinal}), we find first the expansion in FNC of 
\bea
g^{1/4}{\cal D}^2(g^{-1/4}\bar G)= \delta^{ab}\partial_a\partial_b \bar G+{1\over 2}H\bar G+Q^A{}_{b}\,x^b\partial_A\bar G \ ,
\eea
where we defined
\bea
Q^0{}_b=-{1\over 6}R^0{}_b\ ,\quad Q^a{}_b=R^{0a}{}_{0b}\ ,\quad H=\delta^{ab}H_{ab}\ .
\eea
By recursion, it is easy to show that
\bea\non
g^{1/4}{\cal D}^{2n}(g^{-1/4}\bar G)&=&\left[\partial^{2n}+{n\over 2}H\partial^{2(n-1)}+nQ^A{}_bx^b\,\partial_A\partial^{2(n-1)}+n(n-1)Q^A{}_b\,\partial_A\partial^b\partial^{2(n-2)}\right]\bar G\ ,\\
\eea
where we use the convention $\partial^2\bar G=\delta^{ab}\partial_a\partial_b\bar G$, and $\partial^k\bar G=\bar G$ for $k=0\ $. With Eqs.\ (\ref{Hdet}), (\ref{theta}), and (\ref{chris}) we find the expansion of the remaining terms, and the expansion of Eq.\ (\ref{Greeneq}) in FNC and up to two derivatives of the metric finally reads
\bea\label{fulleq}\non
&&\sum_{n=1}^{\infty}\alpha_{2n}\left[ \partial^{2n}\bar G+{n\over 2}H\partial^{2(n-1)}\bar G+nQ^A{}_bx^b\,\partial_A\partial^{2(n-1)}\bar G+
n(n-1)Q^A{}_b\,\partial_A\partial^b\partial^{2(n-2)}\bar G   \right]+\\\non\\
&-&\left(1+Q_{ab}\,x^ax^b\right)\partial_0\partial_0\bar G-Q^a{}_b\,x^b\,\partial_a\bar G+3Q^0{}_c\,x^c\partial_0\bar G-m^2\bar G=-\delta(\tau)\delta(x^a)\ .
\eea
The complexity of this equation can be greatly reduced by performing the $N$-dimensional Fourier transform
\bea\label{fourier}
\bar G(x,x')=\int{d^Nk\over(2\pi)^N}\,e^{ig_{AB}k^{A}x^{B}}\tilde G(p_0^2,p^2)\ ,
\eea
where $k^{A}=(p_0,\vec p)$ is the $N$-momentum, and $x^{B}=(\tau,\vec x)$ are the FNC. The arguments of $\tilde G$ stress the rotational invariance assumed before, together with time-reversal symmetry. As discussed in Appendix \ref{AppB}, the geometrical coefficients are considered $\tau$-independent at any given order, thus the Fourier transform can be calculated to give, after some algebra,
\bea\label{expansion}\non
&&\left(S+p_0^2-m^2\right)\tilde G+\left[ -{1\over 2}H\tilde G+Q^a{}_a\tilde G+Q^A{}_b\,p_A\tilde \partial^b\tilde G \right] DS+Q^A{}_b\,p_A\,p^b\tilde GD^2S+\\\non\\
&&-Q_{ab}\,p_0^2\,\tilde\partial^a\tilde\partial^b\tilde G+Q^a{}_a\tilde G+Q^a{}_b\,p_a\tilde\partial^b\tilde G-3Q^0{}_b\,p_0\tilde\partial^b\tilde G=-1\ .
\eea
In this equation, we have defined 
\bea
S=\sum_{n=1}^{\infty}\alpha_{2n}(-1)^np^{2n}\ ,\quad \tilde\partial^a\tilde G={\partial\tilde G\over \partial p_a}\ ,\quad DS={\partial S\over \partial p^2}\ .
\eea
To solve this equation, we proceed by iteration, similarly as in \cite{BP}. We first write $\tilde G=\tilde G_0+\tilde G_2$, where
\bea\label{G0}
\tilde G_0={1\over m^2-S-p_0^2}\ ,
\eea
is the propagator in flat space. Then, we replace $\tilde G$ with $\tilde G_0+\tilde G_2$ in Eq.\ (\ref{expansion}) and we keep only second order terms. Finally, by using the definition (\ref{G0}) to write $S$ in function of $\tilde G_0$, we find
\bea\label{G2}\non
\tilde G_2&=&-{1\over 2}HD\tilde G_0+Q^0{}_b\,p_0\,p^b D(D-3D_0)\tilde G_0+Q^a{}_a\left[(D+D_0)\tilde G_0-p_0^2DD_0\tilde G_0  \right]+\\\non\\
&+&Q_{ab}\,p^a\,p^b\left[D(D+D_0)\tilde G_0-4\,p_0^2\tilde G_0 D^2\tilde G_0  \right]\ ,
\eea
where $D_0\tilde G=\partial \tilde G/\partial p_0^2$. This is our main result, and it displays the corrections to the flat propagator, $\tilde G_0$, in momentum-space, in terms of geometrical coefficients depending on the background, and operators acting on $\tilde G_0$. These coefficients contain up to two derivatives of the metric. Note that, in this form,  the equation is independent of the dispersion relation adopted. When the dispersion is linear, $S=-p^2$ and $D\tilde G_0=-D_0\tilde G_0=-\tilde G_0^2$. Thus, the above equation reduces to
\bea\label{Grel}
\tilde G_2^{(\rm{rel})}={1\over 2}H\tilde G_0^2+2Q^a{}_a\,p_0^2\tilde G_0^3+8Q_{0b}\,p^0p^b\tilde G_0^3-8Q_{ab}\,p^ap^bp_0^2\tilde G_0^4\ .
\eea

\section{Discussion}

\noindent We now discuss our results, beginning from Eq. (\ref{Grel}). This equation should be contrasted with the well-know Bunch and Parker expansion \cite{BP}, which, to second order, just reads
\bea
\tilde G_2^{(\rm{rel})}={1\over 6}R\tilde G_0^2\ ,\quad \tilde G_0=(p^2-p_0^2+m^2)^{-1}\ ,
\eea
where $R$ is the Ricci scalar. Our expression seems much more complicate, and the reason is due to the choice of FNC instead of RNC. Thus, to transform  Eq.\ (\ref{Grel}) into the expression above, we should go back to Eq.\ (\ref{KG}) and expand in RNC, together with imposing Lorentz-invariance instead of rotational invariance, as we did with Eqs.\ (\ref{symmetries}). In opposition, by using FNC, even with Lorentz-invariant dispersion, the expansion is obtained around the Minkowski metric \emph{and}  a preferred direction with $u^{\mu}_0=\delta^{\mu}_0$ at the lowest order. However, by integrating both expansions in $p_0$, one can see that, in the coincidence limit, the divergent contributions in Eq.\ (\ref{Grel}) add up and coincide with the divergent part of the Bunch and Parker expansion \footnote{Thanks to D.\ Lopez Nacir and D.\ Mazzitelli for pointing out this.}. Therefore, the fundamental Hadamard structure of the two propagators is the same

For Lorentz-breaking dispersions, we have found Eq. (\ref{G2}). The geometrical coefficients depends on various components of the Riemann tensor. Some of these terms become trivial if one assumes that the background has some symmetry. By using the usual transformation law of tensors under coordinate change, $R_{abcd}=R_{\alpha\beta\gamma\delta}e^{\alpha}{}_a\,e^{\beta}{}_b\,e^{\gamma}{}_c\,e^{\delta}{}_d$, it is immediate to show that, in the case of ultra-static metrics, only the first term survives, and we recover the results of Ref.\ \cite{io}.

In the case of a  Bianchi I cosmology, the term  $Q_{0b}$ vanishes, while the remaining ones, when transformed into comoving coordinates, give rise to terms proportional to $R$ and $R_{\mu\nu}u^{\mu}u^{\nu}$. These terms were also found via adiabatic regularization in Ref.\ \cite{mazz2}, however a detailed comparison involves the calculation of the inverse Fourier transform of Eq.\ (\ref{G2}). This appears quite difficult to achieve. In fact, as we showed in Ref. \cite{io}, even in the simple case of a quartic dispersion and ultra-static metric, a coordinate space representation of the Green's function is hard to find by inverse Fourier transform of the momentum-space representation. 

The future plan is to use our results to regularize the UV divergences of the two-point function (and of the stress-energy tensor) with MDR, by operating solely in the momentum-space. This procedure is essential to renormalize the stress tensor in the case of static space-times, such as black holes. For these backgrounds, the Klein-Gordon operator contains higher-order spatial derivatives and WKB methods do not apply, in contrast to the cosmological case. Work in this direction is in progress. 

Finally, we remark that our calculations can be extended to more general cases. For instance, the coefficients of Eq.\ (\ref{G2}) contain up to two derivatives of the metric. Higher order expansions are of course possible, although they are likely to be algebraically rather complicate. Another extension can be obtained by considering a non-geodesics vector field. In this case, one can make use of the generalized Fermi coordinate together with the Fermi-Walker transport, which replaces the usual parallel transport and keeps in account the acceleration of the local frame \cite{poisson-liv,nesterov}. 

\acknowledgments

\noindent I wish to thank R.\ Balbinot, T.\ Jacobson, R.\ Parentani, P.\ Anderson, J.\ Navarro-Salas, and A.\ Ottewill for helpful discussions. I am grateful also to D.\ Lopez Nacir and D.\ Mazzitelli for commenting the first version of the manuscript.

\appendix
\section{Fermi Normal Coordinates }

\begin{figure}[ht]
\includegraphics[width=80mm,height=60mm]{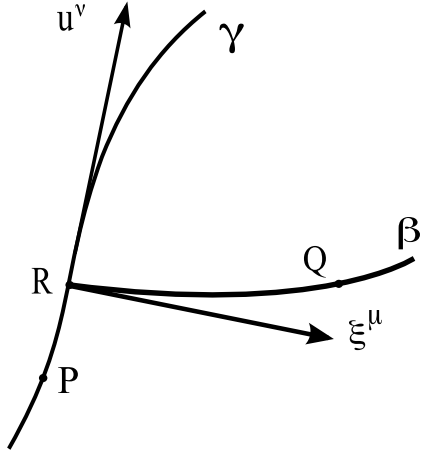} 
\caption{\label{pic} Construction of Fermi coordinates.} 
\end{figure}

\noindent In this appendix, we briefly review the construction of the Fermi Normal Coordinates (FNC). More details can be found in Refs.\ \cite{Misner}-\cite{nesterov}. Let $\gamma$ denote  the integral curve of the vector $u^{\mu}$, which corresponds to the velocity of the observer. We assume that the latter is free-falling, so that $\gamma$ is geodesics. The curve is parameterised by $\tau$, the observer's proper time. Let $P$ be a reference point on $\gamma$, where we set $\tau_P=0$, and $Q$ a point off the curve. The latter is chosen to be close enough to $\gamma$, so that there exist a unique geodesics $\beta$ orthogonal to $\gamma$ an passing through $Q$. We denote by $R$ the point of intersection between $\gamma$ and $\beta$. This point is located at $\tau=\tau_R$. Finally, the vector tangent to $\beta$ at $R$ is called $\xi^{\mu}$, see Fig.\ (\ref{pic}).
At the point $P$ we construct an orthonormal tetrad $\{e^{\mu}_{A}\}$ such that $e^{\mu}_{0}\equiv u^{\mu}$. We use the convention such that frame indices are labeled by $A=0,1,2,\ldots$ or $a=1,2,\ldots$, while general space-time indices are labeled by Greek letters. Since $\gamma$ is a geodesics, we can construct the same tetrad anywhere on the curve by parallel transport. Thus, at any point on the curve, the components of the tetrad satisfy
\bea\label{tetradrelations}
g_{\mu\nu}u^{\mu}u^{\nu}=-1\ ,\quad g_{\mu\nu}\, e^{\mu}_a\, u^{\nu}=0\ ,\quad g_{\mu\nu}\, e^{\mu}_a\, e^{\nu}_b=\delta_{ab}\ .
\eea
The dual tetrad components are defined by
\bea\label{dualtetrad}
e_{\mu}^0=-u_{\mu}\ ,\quad e^a_{\mu}=\delta^{ab}g_{\mu\nu}e^{\mu}_b\ .
\eea
With these, we have the following completeness relations
\bea\label{completeness}
g^{\mu\nu}=-u^{\mu}u^{\nu}+\delta^{ab}\,e_a^{\mu}\,e_b^{\nu}\ ,\quad g_{\mu\nu}=-e_{\mu}^0\,e_{\nu}^0+\delta_{ab}\,e^a_{\mu}\,e^b_{\nu}\ .
\eea
The FNC coordinates of the point $Q$ are defined as
\bea
\tau_Q\equiv \tau_R\ ,\qquad x^a_Q=-e^a_{\alpha}(R)\sigma^{\alpha}(Q,R)\ ,
\eea
where $\sigma^{\alpha}(Q,R)\equiv -\xi^{\alpha}$ is the gradient of the Synge world function relative to $\beta$. By using the completeness relations, one can show that 
\bea
\delta_{ab}\,x^ax^b=2\sigma(Q,R)\equiv s^2\ ,
\eea
where $s$ is the spatial distance between the points $Q$ and $R$ along the geodesic $\beta$. By expanding the function $\sigma$ in $s$, one finds the local metric around $\gamma$, which has the form
\bea\label{FNCmetric}
ds^2=\left(\eta_{AB}+h_{AB}\right)dx^{A}dx^{B}\ ,
\eea
where $\eta_{AB}$ is the Minkowski metric and
\bea\label{h00}
h_{00}&=&-R_{0c0d}\,x^cx^d+{\cal O}(s^3)\ ,\\\non\\\label{h0a}
h_{0a}&=&-{2\over 3}R_{0cad}\,x^cx^d+{\cal O}(s^3)\ ,\\\non\\\label{hab}
h_{ab}&=&-{1\over 3}R_{acbd}\,x^cx^d+{\cal O}(s^3)\ .
\eea
These expressions represent the expansion to second order of the metric tensor around the curve $\gamma$ \footnote{In this context, the order coincides with the number of derivatives of the metric appearing in the coefficients. Thus, as the coefficients are components of the Riemann tensor, which contains two derivatives of the metric, the order is two.}. Note that indices are raised or lowered with the frame Minkowski metric $\eta_{AB}$. The components of the Riemann tensor appearing in these expressions  depend in general on $\tau_R$, as they are evaluated at the point $R$.  If the point $Q$ is on the curve $\gamma$, the above metric reduces to the Minkowski one.
The determinant of the metric can be computed by the identity $\det (I+A)\simeq 1+{\rm Tr} (A) $, as the tensor $h_{AB}$ can be considered as a small correction to the Minkowski metric. Thus, the modulus of the determinant reads
\bea\label{Hdet}
\det |g_{AB}|\equiv g=1-\left(R^{0}{}_{c0d}+{1\over 3}R^a{}_{cad}\right)x^cx^d+{\cal O}(s^3)\equiv 1-H_{cd}\,x^cx^d\ .
\eea
In the theory discussed in this paper,  besides the metric tensor,  there is also the vector field $u^{\mu}$. By construction, this vector is unit time-like and it is the component of the tetrad tangent $\gamma$. However, when the tetrad is parallel-transported along $\beta$ to a point off the curve, the vector $u^{\mu}$ shows non-trivial components along the space-like directions. Suppose that these can be expanded in a Taylor series around the point $R$ as
\bea\label{utaylor}
u^A=u^A|_R+\partial_b u^A|_R\, x^b+{1\over 2}\partial_c\partial_b u^A|_R\, x^bx^c+\ldots\ ,
\eea
with $u^A|_R=\delta^A_0$. Then, parallel transport along $\beta$ implies $\xi^b\nabla_bu^A=0$ for all $\xi^b$. As the connection coefficients vanish along $\gamma$, in opposition to their derivatives, it follows that
\bea
\partial_b u^A|_R&=&-\Gamma^A{}_{bD}u^D|_R=0\ ,\\\non\\
\partial_c\partial_bu^A|_R&=&-\partial_c\Gamma^A{}_{bD}u^D|_R=-R^A{}_{bc0}\ ,
\eea
as it results from the expansion of the connection coefficients, which can be found directly with the help of Eqs.\ (\ref{h00})-(\ref{hab}), and read
\bea\label{chris}
\Gamma^{A}_{ab}=-{2\over 3}\, R^A{}_{(ab)c}\,x^c\ ,\quad \Gamma^A_{0B}=\Gamma^A_{B0}=R^{A}{}_{Bc0}\,x^c\ .
\eea
Thus, the second term in the expansion (\ref{utaylor}) vanishes, and we find
\bea\label{uexp}
u^A=\delta_0^A-{1\over 2}R^A{}_{ij0}\,x^i\,x^j+{\cal O}(s^3)\ .
\eea
With these components we can finally calculate the expansion, to second order, of the induced metric, which reads
\bea\label{indmetric}
q^{AB}=\eta^{AB}+\delta^A_0\delta^B_0-h^{AB}-\delta_0{}^{(A}R^{\,B)}{}_{ij0}\,x^i\,x^j+{\cal O}(s^3)\ .
\eea

\section{Differential Operators in FNC }\label{AppB}

\noindent In this appendix, we find the expansion in FNC of the differential operator  ${\cal D}^2$, which appears in Eq.\ (\ref{KG}). The quickest way is to use the definition \cite{lubo}
\bea
{\cal D}^2f=q^{AB}\nabla_A\nabla_B f+Ku^A\nabla_A f\ ,
\eea
where $K$ is the trace of the intrinsic curvature of the hypersurface orthogonal to $u^{\mu}$. As $K_{AB}=q^C{}_A\,q^D{}_B\nabla _Cu_D$, with the results of the previous appendix, we find that the only non-vanishing terms are
\bea\label{extr}
K_{ab}={1\over 2}R_{0bac}\,x^c\ .
\eea
We notice immediately that the extrinsic curvature is not symmetric. This is a result of the fact that the vector $u^A$ is hypersurface orthogonal only at the point $R$, see Fig.\ (\ref{pic}). However, when we parallel-transport the Fermi tetrad to the point $Q$ along $\beta$, orthogonality is lost, and this gives rise to a non-zero vorticity at $Q$, which, in turn, renders $K_{AB}$ not symmetric \cite{Wald}.  For our calculations however, we just need the trace of $K_{AB}$, which is given by
\bea\label{theta}
K=q^{AB}\nabla_Au_B=-{1\over 2}R^a{}_{0ac}\,x^c=-{1\over 2}R_{0c}\,x^c\ .
\eea
By putting together these results, we finally find
\bea\label{Dfinal}
{\cal D}^2 f=\delta^{ab}\partial_a\partial_b f-{1\over 3}R^{ab}{}_{ac}\,x^c\partial_bf-{1\over 6}R^0{}_{a}\,x^a\partial_0 f\ .
\eea
In order to arrive to this expression, we made use of the symmetries of the Riemann tensor and the rotational invariance of the propagator.  In fact, if $f$ is rotationally invariant (i.e. it depends on powers of $x^2$), we have
\bea\label{symmetries}
h^{0b}\partial_0\partial_b f=-{2\over 3}R^0{}_c{}^b{}_d\,x^c\,x^d\partial_0\partial_b f=0\ .
\eea
and
\bea
h^{ab}\partial_a\partial_b f+{1\over 3}R^{ad}{}_{ac}\,x^c\,\partial_d f=0\ .
\eea
We conclude this appendix with an important remark. We mentioned before that the coefficients of the expansion of $g_{AB}$ depends on $\tau$. Therefore, when we compute terms like $g^{1/4}{\cal D}^2(g^{-1/4} f)$, as in Sec.\ III, we need to know, in principle,  the explicit form of the $\tau$-dependence of the Riemann tensor. In fact, there are $\tau$ derivatives acting on $g$, defined by Eq.\ (\ref{Hdet}). Since we are interested in the short-distance behaviour of the Green's functions, we can take $Q$ close to $P$, which in turn leads to $\tau_R$ close to $\tau_P$, see Fig.\ (\ref{pic}). Hence, we can expand in Taylor series the coefficients $R_{0c0d}$, $R_{0cab}$, and $R_{acdb}$ with respect to $\tau$. However, the results would be coefficients containing more than two derivatives of the metric. Thus, in general, if we fix the order of the expansion to $n$, say, then any $\tau$-derivative of the coefficients yields terms of order $n+1$ or higher, which can be neglected. This means that the coefficients appearing in Eqs.\ (\ref{h00})-(\ref{hab}) can be consistently considered as constant in $\tau$ when hit by a $\tau$-derivative. A further consistency proof follows from the fact that the operator $\square$ can be defined as
\bea
\square f=g^{-1/2}\partial_A\left(g^{1/2}g^{AB}\partial_B f\right)\ ,
\eea
where $\tau$-derivatives of $g$ explicitly appear, or as
\bea
\square f=g^{AB}\nabla_A\nabla_B f\ .
\eea
The two expressions coincide if, and only if, the coefficients are treated as time-independent. For completeness, we report the expression of $\square f$, which reads
\bea\label{fullbox}
\square f=\left(\eta^{AB}-h^{AB} \right)\partial_A\partial_B f-{2\over 3}\,R^{Ab}{}_{db}\,x^d\,\partial_{A}f-R^{0b}{}_{0d}\,x^d\,\partial_b f\ .
\eea

\end{document}